\documentclass[12pt,a4paper]{article} 

\makeatletter
\@addtoreset{equation}{section}
\makeatother

\font\ticp=cmcsc10

\font\cmss=cmss10  
\font\cmsss=cmss10 at 7pt

\def\IZ{\relax\ifmmode\mathchoice 
{\hbox{\cmss Z\kern-.4em Z}}{\hbox{\cmss Z\kern-.4em Z}} 
{\lower.9pt\hbox{\cmsss Z\kern-.4em Z}} 
{\lower1.2pt\hbox{\cmsss Z\kern-.4em Z}}\else{\cmss Z\kern-.4em }\fi} 
\def\IC{\relax\hbox{$\inbar\kern-.3em{\rm C}$}} 
\def\IR{\relax{\rm I\kern-.18em R}}

\def\[{\left [} 
\def\]{\right ]} 
\def\({\left (} 
\def\){\right )}

\def\pa{{\partial}} 
\def\mm{{\rm m}} 
\def\Tr{{\rm Tr}\, } 
 
\def\be{\begin{equation}}       \def\eq{\begin{equation}} 
\def\ee{\end{equation}}         \def\eqe{\end{equation}} 
 
\def\bea{\begin{eqnarray}}      \def\eqa{\begin{eqnarray}} 
\def\ena{\end{eqnarray}}        \def\eea{\end{eqnarray}} 
                                \def\eqae{\end{eqnarray}}

\def\a{\alpha}

\def\d{\delta} 
\def\e{\epsilon}           % Also, \varepsilon 
\def\f{\phi}               %      \varphi 

\def\k{\kappa}                    % Also, \varkappa (see below) 
 
\def\m{\mu} 
\def\n{\nu} 
\def\o{\omega}   
\def\p{\pi}                % Also, \varpi 
\def\r{\rho}                                     %     \varrho 
\def\s{\sigma}                                   %     \varsigma 
\def\t{\tau}

\def\D{\Delta} 
\def\F{\Phi} 
\def\G{\Gamma}

\def\bop#1{\setbox0=\hbox{$#1M$}\mkern1.5mu 
        \vbox{\hrule height0pt depth.04\ht0 
        \hbox{\vrule width.04\ht0 height.9\ht0 \kern.9\ht0 
        \vrule width.04\ht0}\hrule height.04\ht0}\mkern1.5mu} 
\def\Box{{\mathpalette\bop{}}}                        % box 
\def\pa{\partial}                              % curly d 
\def\>{\rangle} %right angle 
 
\def\<{\langle} %left angle 
\def\Dsl{D \hskip-.6em \raise1pt\hbox{$ / $ } } 
 
\def\leftrightarrowfill{$\mathsurround=0pt \mathord\leftarrow \mkern-6mu 
       \cleaders\hbox{$\mkern-2mu \mathord- \mkern-2mu$}\hfill 
       \mkern-6mu \mathord\rightarrow$} 
\def\dvec#1{\vbox{\ialign{##\crcr 
       \leftrightarrowfill\crcr\noalign{\kern-1pt\nointerlineskip} 
       $\hfil\displaystyle{#1}\hfil$\crcr}}}          % <--> accent 
\def\hook#1{{\vrule height#1pt width0.4pt depth0pt}} 
\def\leftrighthookfill#1{$\mathsurround=0pt \mathord\hook#1 
       \hrulefill\mathord\hook#1$} 
\def\underhook#1{\vtop{\ialign{##\crcr                 % |_| under 
       $\hfil\displaystyle{#1}\hfil$\crcr 
       \noalign{\kern-1pt\nointerlineskip\vskip2pt} 
       \leftrighthookfill5\crcr}}} 
\def\smallunderhook#1{\vtop{\ialign{##\crcr      % " for su'scripts 
       $\hfil\scriptstyle{#1}\hfil$\crcr 
       \noalign{\kern-1pt\nointerlineskip\vskip2pt} 
       \leftrighthookfill3\crcr}}} 
\def\sfrac#1#2{{\vphantom1\smash{\lower.5ex\hbox{\small$#1$}}\over 
       \vphantom1\smash{\raise.4ex\hbox{\small$#2$}}}} % alt. fraction 
\def\bfrac#1#2{{\vphantom1\smash{\lower.5ex\hbox{$#1$}}\over 
       \vphantom1\smash{\raise.3ex\hbox{$#2$}}}}      % " 
\def\afrac#1#2{{\vphantom1\smash{\lower.5ex\hbox{$#1$}}\over#2}}  %" 
\def\on#1#2{{\buildrel{\mkern2.5mu#1\mkern-2.5mu}\over{#2}}}%acc.over 
\def\ddt#1{\on{\hbox{\LARGE .\kern-2pt.}}#1}             % double dot 
\def\tdt#1{\on{\hbox{\LARGE .\kern-2pt.\kern-2pt.}}#1}   % triple dot 

\def\to{\rightarrow} 
\def\pa{\partial}

\def\nonu{\nonumber \\{}} 
\def\half{{1 \over 2}}

\def\ppten{{$PP_{10}$ }} 
\def\ppsix{{$PP_{6}\times R^4$ }}

\begin{document} 
 
\begin{flushright} 
hep-th/0309039\\ 
{\ticp TIFR/TH/03-16}\\ 
\end{flushright} 
\vskip .6cm 
\centerline{\Large Comments on D-brane Interactions in PP-wave Backgrounds} 
\medskip 
\vspace*{4.0ex} 
\centerline{{\ticp ATISH DABHOLKAR and JORIS RAEYMAEKERS}} 
\vskip.1in 
\centerline{\it Department of Theoretical Physics} 
\centerline{\it Tata Institute of Fundamental  Research} 
\centerline{\it Homi Bhabha Road, Mumbai, India 400005.} 
\centerline{Email: atish, joris@theory.tifr.res.in} 
\vskip .1in 
\bigskip 
\centerline{ABSTRACT} 
\medskip 

We calculate the interaction potential between widely separated
D-branes in PP-wave backgrounds in string theory as well as in
low-energy supergravity. Timelike and spacelike orientations are
qualitatively different but in both cases the effective brane tensions
and RR charges take the same values as in Minkowski space in
accordance with the expectations from the sigma model perturbation
theory.

\pagebreak 
 
\section{Introduction} 
 
In this note, we study the charges and tensions of Dirichlet branes
and orientifold planes in plane wave backgrounds \cite{blau,
berenstein, penrose, guven}. In a general curved spacetime, the
effective brane tension that is measured from the interaction energy
of two widely separated branes is expected to receive $\a'$
corrections.  {}From the point of view of sigma model perturbation
theory, these corrections will be governed by the $\alpha'$
corrections to the low energy DBI action and will be given in terms of
invariants constructed from the background curvature, field strengths,
and the geometric data of the D-brane embedding
\cite{bachas}. Typically, one would also expect corrections that are 
nonperturbative in $\alpha'$. {}From the point of view of the boundary
conformal field theory, the tension of a D-brane is related to the
regularized dimension of the state space of the CFT
\cite{harvey}. When formulating string theory in a curved background, 
it is an important consistency check whether the Dirac quantization
condition\cite{dnt} for the RR charges is satisfied (see
\cite{bachas2} for a discussion in the case of branes in group
manifolds). The pp-wave background provides another simple example of
a background with a nontrivial metric and Ramond-Ramond fields where
exact worldsheet computation of the D-brane interaction energy is possible.

We find that the brane tensions for half-BPS D-branes in pp-wave
backgrounds are identical to their values in Minkowski space. This
nonrenormalization is in accordance with the expectation based on the
symmetry and the geometry of the plane wave background but the reasons
are different for `time-like' branes that are longitudinal to the
light-cone directions $x^+,\ x^-$ and for `space-like' branes that are
transverse to the light-cone directions.

Time-like branes have translation invariance along $x^-$ which implies
that the D-brane interacts only with those closed string states that
have vanishing $p_-$. For these modes, the metric reduces to the
Minkowski metric and the background appears flat.  Note that this
holds for the full interaction potential and not only for widely
separated branes. In other words, the one point functions of all
closed string modes and not just the massless ones are the same as in
Minkowski space. The same result holds for orientifold planes and
their interactions with D-branes. It follows that for general
orientifolds of the pp-wave background, the orientifold gauge group is
the same as in flat space (a particular case was worked out in
\cite{nemani}).

Space-like branes are not translationally invariant along the
light-cone directions. In this case, the nonrenormalization follows
instead from the special properties of the pp-wave geometry and the
fact that the half-BPS branes that we consider here are totally
geodesic \cite{kostas}.  In the pp-wave background, all local
coordinate invariants constructed out of the background fields vanish.
This is essentially because the only nonvanishing components of the
background fields have a lower $_+$ index and there is no $g^{++}$ to
contract them.  Furthermore, for embeddings that are totally geodesic,
the second fundamental form vanishes. Using the Gauss-Codazzi
equations one can then conclude that all local invariants constructed
using the background fields and the embedding geometry also
vanish. Hence all $\a'$ corrections to charge and tension are expected
to vanish in this background. This can be checked explicitly to
leading order in $\alpha'$ using the corrections to the DBI action
worked out in \cite{dasgupta, bachas} and is expected to be true to all
orders. Note that this argument depends on supersymmetry somewhat
indirectly and only to the extent that the embedding of the
worldvolume of these branes is required to be totally geodesic in
order to preserve half the supersymmetries. Even if the corrections
vanish to all order in $\alpha'$, there remains the possibility of
corrections that are nonperturbative in $\alpha'$, but the plane wave
geometry is topologically trivial and we do not expect any instanton
corrections. It is neverthless important to verify this expectation by
an explicit worldsheet computation because the pp-wave background is
not a small deformation of Minkowski space in any sense. It is not
asymptotically flat and one cannot smoothly interpolate between flat
space and the pp-wave by varying a parameter. The dimensionful
parameter $\m$ that is often introduced can be absorbed in a
coordinate redefinition and is not a physical parameter of the
background. An exact worldsheet computation is therefore desirable to
compare the brane tensions in these completely different backgrounds.

We compute the interaction potentials between a pair of branes and a
brane and an anti-brane in the pp-wave limit of $AdS_5 \times S_5$
(henceforth denoted by $PP_{10}$) and $AdS_3 \times S_3 \times R^4$
(henceforth denoted by $PP_{6}\times R^4$). Strings moving in these
backgrounds can be quantized in the light-cone Green-Schwarz formalism
\cite{metsaev, metsaev2}. D-branes in these backgrounds have been
constructed in \cite{atish, michishita, billo, kostas2, hikida, gaberdiel2,
kostas, sugrapapers} and aspects of their interactions were discussed
in \cite{bergman, gaberdiel, johnson}. The branes we consider here all
preserve half of the kinematical and half of the dynamical
supersymmetries and can be either spacelike or timelike. We calculate
the contribution from the exchange of masssless supergravity modes
from the low energy supergravity and DBI action and find that it
agrees with the string result to all orders in the parameter $\m$
provided that the charges and tensions take the same values as in
Minkowski space\footnote{In
\cite{gaberdiel}, a calculation to leading order in $\m$ was performed
for the D-instanton in \ppten.}. In agreement with \cite{bergman}, we find that the force between two
parallel spacelike D-branes in $PP_{10}$ does not vanish.  For
spacelike branes in \ppsix however, the brane-brane potential is
zero. This can be understood from the fermionic zero modes in the open
string description.

The computation of interaction energy is of interest also from the
point of view of the dual gauge theory. In the dual description, a
single D-brane corresponds to a defect conformal field theory (dCFT)
\cite{karch, dewolfe, bachas3, erdmenger, constable, berenstein2,
lee}.  The interaction energy between two D-branes is expected to
correspond to the Casimir energy between the two defects. The precise
value of the interaction energy from the string computation thus gives
a prediction for the corresponding quantity in the dual
theory. Factorizing the string cylinder diagram in the closed string
channel gives one point functions of off-shell closed string states
emitted from the D-brane. These correspond to one-point functions of
various ambient operators in the dCFT. It would be interesting to
compare some of these predictions by a gauge theory computation.  For
timelike branes, the vanishing of the one point functions for closed
string states with nonzero $p_-$ corresponds to the vanishing of one
point functions of ambient gauge theory operators with nonzero $J$
charge as a consequence of conservation of $J$ charge in the dCFT. To
compare with  the nonzero tadpoles of offshell gravitons with
vanishing $p_-$ however would require a nontrivial computation in the
dCFT and we leave this problem for future work.

This note is organized as follows. Space-like branes are discussed in
sections $\S$\ref{pp10} and $\S$\ref{pp6} and time-like branes are
discussed in $\S4$. The details of the supergravity calculation of the
massless exchanges are given in  appendix \ref{app}.  The
supergravity calculation requires the knowledge of the exact
propagators for the tensor mode fluctuations in this background which
we derive explicitly in the light-cone gauge.

\section{Spacelike branes in $PP_{10}$} 
\label{pp10} 
The \ppten background is given by (see appendix \ref{pp10modes} for
more details on our conventions):
\bea 
ds^2 &=& 2 d x^+ d x^- - \m^2 x^I x^I (dx^+)^2 + dx^I dx^I\nonu
R_{I++J} &=& - \m^2 \d_{IJ} \qquad R_{++} = 8 \m^2\nonu
F_{+1234} &=& F_{+5678} = 4 \m 
\eea 
where $I = 1 \ldots 8$. The Ramond-Ramond background breaks the
$SO(8)$ acting on the $x^I$ to $SO(4) \times SO'(4)$, the first factor
acting on $x^i,\ i = 1 \ldots 4$ and the second one acting on
$x^{i'},\ i' = 5 \ldots 8$. Denoting a spacelike D-brane with $m$
worldvolume directions along the $x^i$ and $n$ worldvolume directions
along the $x^{i'}$ by $(m,n)$, the branes preserving half of the
kinematical and half of the dynamical supersymmetries are of the type
$(m, m+ 2)$ (or, equivalently, $(m+2, m)$) with $m=0, 1$ or $2$
\cite{atish, kostas}. Hence we are to consider D1, D3 and
D5-branes.  These are to be placed at the origin of the $SO(4)
\times SO'(4)$ directions in order to preserve the aformentioned
supersymmetries. We will calculate the interaction energy between
pairs of (anti-) D-branes separated along the $x^+,\ x^-$
directions.
 
\subsection{String calculation} 
The string theory calculation of the interaction energy between a pair
of D-branes of the same dimension was performed in \cite{bergman}; we
will briefly review it here in order to extract the contribution from
the lowest lying string modes.
 
We would like to perform the string calculation in the open string
loop channel to get a correctly normalized amplitude.  However, in the
standard light-cone gauge $X^+ = p_- \t$, $X^\pm$ are automatically
Neumann directions. We can remedy this by using a nonstandard
light-cone gauge for the open string \cite{bergman, gaberdiel} in
which $X^\pm$ are Dirichlet directions. Here, one quantizes the open
strings stretching from one brane to the other in the gauge $$ X^+ =
{r^+ \over \p} \s $$ where $r^+$ is the brane separation along the
$x^+$ coordinate and $\s$ is the worldsheet coordinate, $\s \in
[0,\p]$. The Virasoro constraints then determine $X^-$ to be a
Dirichlet direction as well\footnote{Note that this gauge is
consistent with the Virasoro constraints only if the worldsheet is
Euclidean \cite{gaberdiel}.}.  In this gauge, the worldsheet action contains eight
massive bosons and fermions with mass $$
\mm = {\m r^+ \over \p}. 
$$     
The interaction energy between branes can be written as 
\bea 
ET &=& 2 \cdot \half  i \Tr (-1)^{F_s} \ln (L_0 - i \e)\nonu 
&=& i \int_0^\infty {ds \over s} \Tr (-1)^{F_s} e^{- i (L_0 - i \e ) s}  
\label{propertime} 
\eea 
where $F_s$ is the spacetime fermion number and the trace is taken  
in the space of open string states stretching between the branes. $L_0$ is the 
generator of worldheet time translations and can be written as 
$L_0 = p_- H^{lc}$ with $H^{lc}$ the light-cone Hamiltonian. In writing 
(\ref{propertime}), we chose to work in Lorentzian signature for spacetime 
 with a suitable $i \e$ prescrition \cite{DeWitt, Birrell}.  
 
For a Dp-brane interacting with an anti-Dp-brane, $L_0$ receives 
a contribution from the separation from the strings being stretched along  
the transverse directions $x^+,\ x^-$  
and contributions from harmonic oscillators with frequencies $\o_k = {\rm sign} (k) \sqrt{k^2 + \mm^2}$, 
where $k$ is integer for the bosonic oscillators and half-integer   for the fermionic ones. 
The resulting interaction energy is  
\be 
E_{Dp-\bar Dp} = i \int_0^\infty {ds e^{-\e s} \over s} e^{-2 \p i s \left({ 2 r^+ r^-  
\over 4 \p^2 \a '}\right)} (2 i \sin \p \mm s)^{3-p}   
\left( 
{f_4^{(\mm)} (q)\over f_1^{(\mm)} (q)}\right)^8, 
\label{ddbarpp10} 
\ee 
where we have defined modified $f$-functions as in \cite{bergman}: 
\bea 
f_1^{(\mm)}(q) & =& q^{-\Delta_\mm} (1-q^\mm)^{{1\over 2}} 
\prod_{n=1}^{\infty} \left(1 - q^{\sqrt{\mm^2+n^2}}\right) \,, 
\label{fdef}\\ 
f_2^{(\mm)}(q) & = &q^{- \Delta_\mm} (1+q^\mm)^{{1\over 2}} 
\prod_{n=1}^{\infty} \left(1 + q^{\sqrt{\mm^2+n^2}}\right) \,, 
\label{f2def}\\ 
f_3^{(\mm)}(q) & =& q^{-\Delta^\prime_\mm} 
\prod_{n=1}^{\infty} \left(1 + q^{\sqrt{\mm^2+(n-1/2)^2}}\right) \,, 
\label{f3def}\\ 
f_4^{(\mm)}(q) & =& q^{-\Delta^\prime_\mm} 
\prod_{n=1}^{\infty} \left(1 - q^{\sqrt{\mm^2+(n-1/2)^2}}\right) \,, 
\label{f4def} 
\eea 
and $\Delta_\mm$ and $\Delta^\prime_\mm$ are defined by 
\bea 
\Delta_\mm & =& -{1\over (2\pi)^2} \sum_{p=1}^{\infty} 
\int_0^\infty ds \, e^{-p^2 s} e^{-\pi^2 \mm^2 / s} \,, \cr 
\Delta^\prime_\mm & =& -{1\over (2\pi)^2} \sum_{p=1}^{\infty} 
(-1)^p \int_0^\infty ds \, e^{-p^2 s} e^{-\pi^2 \mm^2 / s}\,. 
\label{Deltadef} 
\eea 
For two parallel Dp-branes, the harmonic oscillator frequencies 
are $\o_k$ with $k$ integer for both the bosons and fermions. 
The fermionic ``zero-modes'' have frequency $\mm$ and give a  
nonzero contribution to the interaction energy. The result is  
\be 
E_{Dp- Dp} = i \int_0^\infty {ds e^{-\e s}\over s} e^{-2 \p i s \left({ 2 r^+ r^-  
\over 4 \p^2 \a '}\right)} (2 i\sin \p \mm s)^{3-p}   
\label{ddpp10} 
\ee 
 
The large distance behaviour of (\ref{ddbarpp10}) and (\ref{ddpp10})  comes from  
the leading behaviour of the integrand for small $s$. This can be  
extracted using the modular transformations  
\be 
f_1^{(\mm)}(s)  = f_1^{(\widehat{\mm})}(-1/s) \,, \qquad 
f_2^{(\mm)}(s)  = f_4^{(\widehat{\mm})}(-1/s) \,,\qquad 
f_3^{(\mm)}(s)  = f_3^{(\widehat{\mm})}(-1/s), 
\label{beautiful} 
\ee 
where 
\be 
\widehat{\mm}= i \mm\, s\,. 
\label{mhat} 
\ee 
This gives the leading behaviour 
\bea 
E_{Dp-\bar Dp} &=& - (4 \p^2 \a ')^{3 - p}  
 (2\p r^- )^{p-3} \m^{3-p} \cot^4 \m r^+   \G(3-p)+ \ldots \nonu 
E_{Dp-Dp} &=& - (4 \p^2 \a ')^{3 - p}  
 (2\p r^- )^{p-3} \m^{3-p}    \G(3-p)+ \ldots 
\eea  
The expression in the first line diverges in the flat space limit $\m \to 0$; 
this is the standard divergence due to the infinite volume of the brane. 
We can separate out the volume factor by rewriting the result in terms 
of the integrated propagator $I_0^{9-p}(r^+,r^-)$ over the worldvolume 
directions with the remaining transverse pp-wave coordinates 
set to zero (see in (\ref{intproppp10})): 
\bea 
E_{Dp-\bar Dp} &=&  -4 \p (4 \p^2 \a ')^{3 - p}\cos^4 \m r^+ 
I_0^{9-p} ( r^+, r^-)+ \ldots\nonu 
E_{Dp-Dp} &=&  - 4 \p (4 \p^2 \a ')^{3 - p}\sin^4 \m r^+ 
I_0^{9-p} ( r^+, r^-)+ \ldots. 
\label{potspp10} 
\eea 
One recovers the correct flat space expression \cite{Pol} as $\m \to 0$ taking 
into account that, from (\ref{momprop}), 
$$ 
\lim_{\m \to 0} I_0^{9-p} ( r^+, r^-) = 
V_{p+1} G_0^{9-p} ( r^+, r^-) 
$$ 
where $V_{p+1}$ is the divergent D-brane volume and  
$G_0^{9-p} ( r^+, r^-)$ stands for the Minkowski 
space scalar propagator integrated over the worldvolume 
directions.

\subsection{Field theory calculation} 
 
We will presently see how the long-range 
potentials (\ref{potspp10}) are 
reproduced exactly from  
the type IIB supergravity action (\ref{IIBaction})  
supplemented with 
D-brane source terms  
\be 
S_p =  -T_p \int d^{p+1} x   
 \sqrt{- \tilde g} e^{{p-3 \over 4} \Phi}  +  \m_p \int A_{[p+1]}  
\label{sourceaction} 
\ee  
where $\tilde g$ stands for the induced metric  
on the worldvolume and $T_p$ and $\m_p$ are the brane tension and RR 
charge respectively. In appendix \ref{pp10modes} we expand the  
bulk action to quadratic order in the fluctuations around 
the \ppten background, adopting the light-cone gauge for 
the fluctuations. The resulting action is a sum of decoupled terms 
characterized by an integer $c$: 
\be 
S_\psi = {1 \over 4 \k^2} \int d^{10}x  \psi^\dagger ( \Box - 2 i \mu c \pa_-) \psi. 
\label{psiaction} 
\ee 
In general, $\psi$ is in a tensor representation of $SO(4) \times SO'(4)$ and 
a contraction of tensor indices is understood. 
The decoupled fields $\psi$, their values of $c$ and their 
$SO(4) \times SO'(4)$ representations are 
given in the appendix in table \ref{pp10table}. 
Expanding the source action (\ref{sourceaction}) to linear 
order in the fluctuations we get source terms for the components $\psi_\a$ of the form 
\be 
S_{\rm source} = \int d^{10}x\ \d^{9-p} (x - x_0)\ k ( \psi_\a + \e \bar \psi_\a) 
\label{psisource} 
\ee 
where $k$ is a constant proportional to either $T_p$ or $\m_p$ and $\e
= \pm 1$. The contribution of such a mode to the interaction energy
can then be written in terms of the integrated propagator and the
constants $(c,k,\e)$. For example, if $\psi$ is an $SO(4) \times
SO'(4)$ singlet one gets a contribution to the interaction energy
\be 
E_{(c,k,\e)}  = 8 \e \k^2 k^2  \cos \m c r^+ I_0^{9-p} ( r^+, r^-). 
\label{intencontr} 
\ee 
When $\psi$ is in a tensor representation $SO(4) \times SO'(4)$,  
this expression can get an extra overall factor from the fact that 
one has to use a propagator with the right symmetry properties. 
%one has to correct for 
%the fact that (\ref{psiaction}) is not canonically normalized
%\footnote{Also, if $\psi$ 
%is a traceless tensor, one has to project the propagator on 
%its traceless part}.  
 
\subsubsection{D1-brane} 
 
We can take the worldvolume along the directions $x^1,x^2$. The source
terms are given by
\bea 
{\cal L}_{\rm source} &=& {i T_1 \over 2} (h_{11} + h_{22} - \f ) \pm
\m_1 a_{12}\nonu
&=& {i T_1 \over \sqrt{2}}(h_{11}^\perp + h_{22}^\perp) + {i T_1 \over
4} (H + \bar H) - {i T_1 \over 2} \f + {i T_1 \over 4} h
\pm { \m_1 \over \sqrt{2}}( G_{12} + \bar G_{12}) \nonumber
\eea  
where the upper (lower) sign applies to a brane (antibrane)
source. The factors of $i$ multiplying the tension arise because $-
\tilde g$ is negative for spacelike branes. The trace $h$ doesn't
propagate in the light-cone gauge. The constants $(c,k,\e)$ for the
other source terms are summarized in the following table:
\begin{center} 
\begin{tabular}{|c||c|c|c|c|}\hline 
$\psi$&$ h_{11}^\perp,\; h_{22}^\perp$&$H$& $ \f$&$G_{12}$\\\hline 
$(c,k,\e)$&$(0,{iT_1 \over 2 \sqrt{2}},1)$&$(4,{iT_1\over 4},1)$ 
&$(0,{-i T_1\over 4},1)$& 
$(2,{\pm  \m_1 \over \sqrt{2}},1)$\\\hline 
\end{tabular} 
\label{D1pp10table} 
\end{center} 
Summing up all contributions to the interaction energy gives 
\be 
E = - 4 \k^2  \left[ { T_1^2\over 8} \cos 4 \m r^+ + {3 \over 8}T_1^2 
 \mp {\m_1^2 \over 2} \cos 2 \m r^+ 
 \right] I_0^{8}(r^+, r^-) 
\label{D1pp10} 
\ee 
where the upper (lower) sign applies to the brane-brane
(brane-antibrane) system.
%& = & 4 \k^2 T_1^2 \sin^4 \m r^+ I_0^{8}(r^+, r^-) \qquad 
%{\rm D1-D1\ system}\\ 
%& &4 \k^2 T_1^2 \cos^4 \m r^+ I_0^{8}(r^+, r^-) \qquad 
%{\rm D1-\bar D1\ system} 
%\eea 
%where, in the last equality, 
%we used a trigonometric identity \cite{gradshteyn}. 
 
\subsubsection{D3-brane} 
 
We take the worldvolume directions to be $x^1,x^2,x^3,x^5$. The source
terms are
\bea 
{\cal L}_{\rm source} &=& {i T_3 \over 2} ( h_{11} + h_{22} + h_{33} +
h_{55} ) + \m_3 a_{1235}\nonu &=& {i T_3 \over \sqrt{2}} ( h_{11}^\perp +
h_{22}^\perp + h_{33}^\perp + h_{55}^\perp ) + {i T_3 \over 4} (H +
\bar H) + {1 \over 2} h \mp {i \m_3 \over 2} (H_{45} - \bar H_{45})
\nonumber
\eea 
These give the following interaction energy contributions 
\begin{center} 
\begin{tabular}{|c||c|c|c|}\hline 
$\psi$&$ h_{11}^\perp,\; h_{22}^\perp,\;  h_{33}^\perp,\;  h_{55}^\perp $&$H$& $H_{45}$\\\hline 
$(c,k,\e)$&$(0,{iT_3 \over 2 \sqrt{2}},1)$&$(4,{iT_3\over 4},1)$ 
&$(2,{\mp i \m_3 \over 2},-1)$\\\hline 
\end{tabular} 
\end{center} 
The total is 
\be 
E = - 4 \k^2  \left[ { T_3^2\over 8} \cos 4 \m r^++ {3 T_3^2 \over 8} 
 \mp {\m_3^2 \over 2} \cos 2 \m r^+ 
 \right] I_0^{6}(r^+, r^-) 
\label{D3pp10} 
\ee 
                                
\subsubsection{D5-brane} 
 
The calculation is the same as for the D1-brane due to S-duality 
invariance of the type IIB supergravity action and the fact that 
the \ppten background is also S-duality invariant. 
The end result is again 
\be 
E =- 4 \k^2  \left[ { T_5^2\over 8} \cos 4 \m r^++ {3 T_5^2 \over 8} 
 \mp {\m_5^2 \over 2} \cos 2 \m r^+ 
 \right] I_0^{4}(r^+, r^-) 
\label{D5pp10} 
\ee 
 
\subsubsection{D-brane charges and tensions} 
 
Comparing the string calculation (\ref{potspp10}) with the field
theory results (\ref{D1pp10}, \ref{D3pp10}, \ref{D5pp10}) we find
agreement only if the charges and tensions are equal and their
numerical value is the same as in Minkowski space \cite{Pol}: \be
T_p^2 = \m_p^2 = {\p (4 \p^2 \a')^{(3-p)} \over \k^2} \qquad p=1,3,5.
\ee To see this, one has to use the trigonometric identities \bea
\cos^4 x &=& {1\over 8} \cos 4 x + {3\over 8} + \half \cos 2 x\nonu
\sin^4 x &=& {1\over 8} \cos 4 x + {3\over 8} - \half \cos 2 x.  \eea
These identities in a sense encode the equivalence between the open-
and closed string descriptions and were instrumental in proving
Cardy's condition for boundary states in \cite{bergman}.
 
\section{Spacelike branes in $PP_6\times R^4$} 
\label{pp6} 
Our conventions for the $PP_6\times R^4$ coordinates and background fields 
are (see appendix \ref{pp6modes} for more details)  
\bea 
ds^2 &=& 2 d x^+ d x^- - \m^2 (z \bar z + w \bar w)(dx^+)^2 + dz d\bar z 
+ dw d\bar w + dx^a dx^a\nonu 
R_{z++\bar z} &=& - \half \m^2 \qquad R_{w++\bar w} = - \half \m^2  
 \qquad R_{++} = 4 \m^2\nonu
F_{+z \bar z} &=& F_{+w \bar w} = i \m 
\eea  
The allowed D-branes in $PP_6\times R^4$ were classified  
in \cite{michishita}. Here, we restrict attention to spacelike branes 
with worldvolumes lying in the $PP_6$ part of the geometry.  
Branes with worldvolume directions along the $R^4$ (and their tensions) 
can be obtained by applying T-duality along the $R^4$ directions. 
Denoting by $(m,n)$ a brane with $m$ directions along the $U(1)$ and $n$ 
directions along $U'(1)$, the branes preserving half the kinematical and half the 
dynamical supersymmetries are of the type $(m,m)$ with $m=1,2$.   
This leaves the D1 and D3 branes to be considered. 
Supersymmetry requires that the D1-brane be 
placed at the origin of the transverse $U(1) \times U'(1)$ directions. 
 
\subsection{String calculation} 
The string theory calculation of the interaction energy  
proceeds as in the $PP_{10}$ case. After fixing a non-standard 
light-cone gauge, the worldsheet action for strings stretching 
between branes contains four massive bosons and 
fermions with mass 
$$ 
\mm = {\m r^+ \over \p} 
$$    
as well as four massless bosons and fermions. The interaction 
energy between a Dp-brane and an  
anti-Dp-brane is given by the  
open string one-loop 
amplitude  
\be 
E_{Dp-\bar Dp} =i \int_0^\infty {ds\; e^{-\e s} \over s} e^{-2 \p i s \left({ 2 r^+ r^- +  
r^ar^a \over 4 \p^2 \a '}\right)} (2 i \sin \p \mm s)^{1-p}   
\left( 
{f_4^{(\mm)} (q)\over f_1^{(\mm)} (q)}\right)^4 \left( {f_4^{(0)} (q)  
\over f_1^{(0)} (q)} \right)^4 
\label{ddbarpp6} 
\ee 
The leading contribution comes from massless exchanges and is  
given by 
\be 
E_{Dp-\bar Dp} = - 4 \p (4 \p^2 \a ')^{3 - p}\cos^2 \m r^+ 
I_0^{9-p} ( r^+, r^-, r^a) 
\label{potddbarpp6} 
\ee 
where $I_0^{9-p} ( r^+, r^-, r^a)$ stands for the scalar  
propagator integrated over $p+1$ longitudinal D-brane directions  
with the remaining transverse pp-wave coordinates 
set to zero (see (\ref{intproppp6})).  
We again observe that (\ref{potddbarpp6}) reduces to  
the correct flat-space expression \cite{Pol}  
as $\m \to 0$. 
 
Contrary to the $PP_{10}$ case, the interaction energy  
between two parallel branes in 
$PP_6 \times R^4$ is zero: 
\be 
E_{Dp-Dp} =  0. 
\label{potddpp6} 
\ee 
This follows immediately from the fact that an open 
string stretching between the branes has  
four actual (meaning zero-frequency) fermionic zero modes. 
 
\subsection{Field theory calculation} 
 
In appendix \ref{pp6modes} we expand the bosonic type IIB 
action around the \ppsix background and identify 
the independent fluctuations.  
The field theory calculation of the interaction energy again reduces to a sum of 
contributions of the form (\ref{intencontr}), characterized 
by constants $(c,k,\e)$ which can be read off by writing the D-brane 
 source terms in terms of the decoupled fields listed 
in table \ref{pp6table} of the appendix.

\subsubsection{D1-brane} 
 
We take the worldvolume directions to be $x^1, x^3$. 
Expressing the D-brane sources in terms of the decoupling fields  
in table \ref{pp6table} one gets the 
worldvolume Lagrangian 
\bea 
{\cal L}_{\rm source} &=& {iT_1 \over 2} (h_{11} + h_{33} - \f) \pm  
 \m_1 a_{13}\nonu
&=& {iT_1 \over 4}(\tilde h_{zz} +\tilde h_{\bar z\bar z} +\tilde h_{ww} +\tilde h_{\bar w\bar w} 
 + H + \bar H + H' + \bar H') + {iT_1 \over 4} h \nonu
&\pm&  {\m_1  \over 4} (H^+_{zw} +H^+_{\bar z \bar w} -H^-_{zw}-H^-_{\bar z \bar w}+ 
H^+_{z\bar w} +H^+_{\bar z  w} -H^-_{z\bar w}-H^-_{\bar z w}). \nonumber
\eea 
where the upper sign refers to a brane and the lower one  
to an anti-brane. 
The trace $h$ does not propagate in the light-cone gauge. 
The other fields give contributions 
of the form (\ref{intencontr}) to the interaction energy. These 
are summarized in the following table: 
\begin{center} 
\begin{tabular}{|c||c|c|c|c|c|}\hline 
$\psi$&$\tilde h_{zz},\; \tilde h_{ww}$&$ H ,\; H'$&$H^+_{zw},\; H^-_{zw}$& 
$H^+_{z\bar w}$&$ H^-_{z\bar w}$ \\\hline 
$(c,k,\e)$&$(0,{iT_1 \over 4},1)$&$(2,{iT_1\over 4},1)$ 
&$(0,{\pm  \m_1 \over 4},1)$&$(2,{\pm  \m_1\over 4},1)$ &$(-2,{\pm  \m_1\over 4},1)$\\\hline 
\end{tabular} 
\end{center} 
where the upper sign applies to the brane-brane system and the lower 
sign applies to the brane-antibrane configuration. 
Summing all contributions, one gets the total 
interaction energy 
\bea 
E &=& - 2 \k^2 \left[ T_1^2 (\cos 2 \m r^+ + 1) \mp \m_1^2 (\cos 2 \m r^+ + 1) 
\right] I_0^{8} ( r^+, r^-, r^a)\\ 
&=& - 2 \k^2 [ T_1^2 \mp \m_1^2 ] \cos^2 \m r^+  I_0^{8} ( r^+, r^-, r^a) 
\label{D1pp6} 
\eea

\subsubsection{D3-brane} 
 
The worldvolume directions are $x^1,x^2, x^3,x^4$. 
The source terms are 
\bea 
{\cal L}_{\rm source} &=& {iT_3 \over 2} (h_{11} + h_{22}+ h_{33} +h_{44}) \pm  
 \m_3 a_{1234}\nonu
&=& {iT_3 \over 4}(H + \bar H + H' + \bar H')+ {iT_3 \over \sqrt{2}} H_0 + 
{iT_3 \over 2} h \nonu 
&\pm&  {\m_3  \over 2 \sqrt{2} } (G + \bar G) \pm  {\m_3  \over \sqrt{2} } G_0 . 
\nonumber
\eea 
The  contributions 
to the interaction energy 
are summarized in the following table: 
\begin{center} 
\begin{tabular}{|c||c|c|c|c|}\hline 
$\psi$&$ H,\; H'$&$H_0$& $ G ,\; G'$&$G_0$\\\hline 
$(c,k,\e)$&$(2,{iT_3 \over 4},1)$&$(0,{i T_3\over 2 \sqrt{2}},1)$ 
&$(2,{\pm  \m_3 \over 2 \sqrt{2}},1)$& 
$(0,{\pm  \m_3\over 2 \sqrt{2}},1)$\\\hline 
\end{tabular} 
\end{center} 
 Summing all contributions, one gets the total 
interaction energy 
\bea 
E &=& -2 \k^2 \left[ T_3^2 (\cos 2 \m r^+ + 1) \mp \m_3^2 (\cos 2 \m r^+ + 1) 
\right] I_0^{6} ( r^+, r^-, r^a)\\ 
&=& -2 \k^2 [ T_3^2 \mp \m_3^2 ] \cos^2 \m r^+  I_0^{6} ( r^+, r^-, r^a) 
\label{D3pp6} 
\eea 
Again, the upper sign applies to the brane-brane system and the lower 
sign applies to the brane-antibrane configuration. 
 
\subsection{D-brane charges and tensions} 
 
Comparing the results (\ref{potddbarpp6}) and (\ref{potddpp6}) of the string 
calculation with the field theory results (\ref{D1pp6}) and (\ref{D3pp6}), 
we find the value of the D-brane charge and tension: 
\be 
T_p^2 = \m_p^2 = {\p (4 \p^2 \a')^{(3-p)} \over \k^2} \qquad p=1,3. 
\ee 
These values are again the same as in Minkowski space.

\section{Timelike Branes}\label{+-branes} 
 
In this section we will argue that interactions between timelike
D-branes that extend along the $x^+,\ x^-$ directions in a plane wave
geometry are the same as in Minkowski geometry. This is a consequence
of the fact that these branes preserve translation invariance in the
$x^-$ direction. Similar results hold for orientifold planes.
 
For definiteness, we illustrate this in the $PP_6 \times R^4$
background and comment on the generalization to other pp-wave
backgrounds at the end of this section. Let us consider a brane-brane
or a brane-anti-brane pair in this background separated along the
$R^4$ directions. From the point of view of the low-energy effective
field theory, the long-range interaction potential comes from the
exchange of massless modes between the branes. The Feynman propagator
$G_c(x_1,x_2)$ of such modes is given by (see (\ref{momproppp6}))
$$
 \sum_{\bf n} \int { dp_+ dp_- d^4 p_a
\over (2 \p)^6} {e^{i \left( 
p_+ (x^+_1 - x^{+}_2) + p_- (x^-_1 - x^{-}_2) + p_a (x^a_1 -  x^a_2) 
\right)} \psi_{\bf n}^{(\m p_-)}(x^A_1)\psi_{\bf n}^{(\m p_-)}(x^A_2) \over  
2 p_+ p_- + \m p_- \sum_I (n_I + \half) + 2 c \m p_- + p_a p_a - i \e}.
$$ 
In calculating the interaction energy between a pair of branes, we
have to integrate (\ref{momprop}) over the worldvolume directions
which include $x^-_1, x^-_2$. This gives a delta function for $p_-$,
hence the result is independent of $\m$. We recover the flat space
result for the interaction energy (times the interaction time):
\be 
ET = 2\k^2 (T_p^2 \mp \m_p^2) G_{9-p}(r^a) V_{p+1}. 
\label{fieldresult1} 
\ee 
A similar argument can be made for the exchange of massive modes,
which shows that the full brane-antibrane interaction potential is the
same as in Minkowski space. In the language of boundary
states\footnote{ The only subtlety here is that one has to quantize
the closed string in a nonstandard light-cone gauge in order for the
coordinates $X^\pm$ to have the right boundary conditions
\cite{bergman}.}, The interaction energy, say, between a brane and an
anti-brane is given by the overlap $\langle D \bar p | \D | Dp
\rangle$ where $\D$ is the closed string propagator and $|Dp\rangle,
|D \bar p\rangle$ are the boundary states. Since the boundary states
satisfy $p_-|Dp\rangle = p_-|D \bar p\rangle =0$, the closed string
propagator is projected on the $p_-=0$ subspace and these states
propagate as in flat space.
 
It is instructive to see how the same conclusion follows from a
calculation in the open string picture. This, in some sense, is the
natural picture to use for timelike branes since the open strings can
be quantized in the usual light-cone gauge $X^+ = p_- \t$. In this
gauge, the coordinates $X^\pm$ automatically obey Neumann boundary
conditions \cite{atish}.  The brane-antibrane interaction energy is
given by
\be 
ET = i V_{+-} \int_0^\infty {ds\; e^{-\e s} \over s} \int {d p_+ dp_-
\over (2 \p)^2} e^{ - 4 \p \a' i s p_+ p_-} Z(s, \m p_-)
\label{stringampl1} 
\ee 
where 
\bea  
Z(s, \m p_-) &=& \Tr (-1)^{F_s} q^{\a' p_- H^{\rm lc}}\nonu &=&
q^{r^ar^a \over (2 \p)^2 \a '} (2 i \sin \p \m p_- s)^{3 - p} \left(
{f_4^{(\m p_-)} (q)f_4^{(0)} (q) \over f_1^{(\m p_-)} (q)f_1^{(0)}
(q)} \right)^4.
\eea 
The function $Z(s, \m p_-)$ is the partition function for a combined
system of four massive scalars and fermions with mass $\m p_-$ and
four massless scalars and fermions, with appropriate boundary
conditions. Returning to (\ref{stringampl1}), we see that the $p_+$
integral yields a delta function. Integrating over $p_-$ we find
\be 
ET  = i V_{+-} \int_0^\infty {ds \; e^{-\e s} \over 8 \p^2 \a' s^2} Z(s , 0).
\ee 
In particular, using a modular transformation to extract the small $s$
behaviour of the integrand, one finds the dominating contribution for
widely separated branes
\be  
ET = 4 \p (4 \p^2 \a')^{(3-p)} V_{p+1} G_{9-p}(r^a)+ \ldots . 
\label{stringresult1} 
\ee 
Comparing with (\ref{fieldresult1}) we find 
\be 
T_p^2 = \m_p^2 = {\p (4 \p^2 \a')^{(3-p)} \over \k^2}, 
\ee 
which is indeed the flat-space value \cite{Pol}. 
 
Similar results hold for the interactions between timelike orientifold
planes, and between orientifold planes and D-branes in \ppsix.  This
can be argued from the fact that the crosscap state is annilated by
$p_-$ or, in the open string picture, from the fact that the
interaction energy is again of the form (\ref{stringampl1}) but with a
different function $Z(s, \m p_-)$
\cite{nemani, hammou}. Hence the tadpole cancellation conditions for timelike 
orientifolds in \ppsix are the same as in Minkowski space. The above
argument shows that not only the massless tadpoles but the one point
functions on a disk of even the massive string modes take the same
value as in Minkowski space.

\leftline{\bf Acknowledgements}

We would like to thank Kostas Skenderis and Marika Taylor for very
useful discussions and comments.
 
\begin{appendix}

\section{Massless modes and propagators in $PP_{6}$ and $PP_{10}$}\label{app} 
In this appendix we obtain the Lagrangian and the propagator for  
the bosonic massless supergravity modes in the pp-wave background. 
The starting point is the bosonic part of the 
type IIB action in the Einstein  
frame: 
\bea 
S &=& {1 \over 2 \k^2} \int d^{10}x \sqrt{-g}  
\left[ R - \half (\pa \F)^2 - {H_{[3]}^2 \over 2\cdot 3!} 
 - \half e^{-2 \F} F_{[1]}^2 - {e^{- \F} \tilde F_{[3]}^2 \over 2\cdot 3!} 
- {\tilde F_{[5]}^2  \over 4\cdot 5!}
\right]\nonumber\\ 
&& - {1 \over 2} \int A_{[4]} \wedge H_{[3]} \wedge F_{[3]} 
\label{IIBaction} 
\eea 
where $F_{[2n + 1]} \equiv d A_{[2n]},\ H_{[3]} \equiv d B_{[2]}$ 
and 
\bea 
\tilde F_{[3]} &\equiv& F_{[3]} - A_{[0]} \wedge H_{[3]}\nonu 
\tilde F_{[5]} &\equiv& F_{[5]} - \half A_{[2]} \wedge H_{[3]} 
+ \half B_{[2]} \wedge F_{[3]} 
\eea 
In the following we will expand the action around the  
$PP_{6}$ and $PP_{10}$ backgrounds to quadratic order 
in the fluctuations 
\bea 
g_{\m \n} &\to& g_{\m \n} + h_{\m \n}\nonu
\F &\to& \F + \f\nonu
B_{[2]} &\to&B_{[2]}+b_{[2]}\nonu 
A_{[2n]} &\to& A_{[2n]}+ a_{[2n]}. 
\label{flucts} 
\eea 
It's convenient to split the metric fluctuations into 
a trace part $h$ and a traceless tensor $h^T_{\m\n}$.  
We adopt the light-cone gauge for the fluctuations: 
\be 
h_{-\m} = b_{-\m} = a_{-\m_1 \ldots \m_{2n-1}}  = 0. 
\ee 
In this gauge, after shifting the fields 
with a $_+$ index, one finds that $h,\ h^T_{+\m},\ b_{+\m}$ and $ 
a_{+\m_1 \ldots \m_{2n-1}}$ decouple. This situation is familiar from 
the light-cone gauge in Minkowski space (see e.g. \cite{Siegel}). 
Hence the only propagating fields are the transverse modes 
$h_{IJ},\  b_{IJ},\ a_{I_1 \ldots I_{2n}};\ I,J,\ldots = 1 \ldots 8$. 
In the presence of general sources, the gauge-fixed Lagrangian  
contains Coulomb-like terms as a result of shifting the fields.  
In the cases we consider, these are absent because spacelike branes 
do not provide a source for the fields with a $_+$ index.

\subsection{Massless modes in $PP_{10}$} 
\label{pp10modes} 
We use the following index conventions: 
\bea 
\m,\n,\ldots = 0, \ldots,9 &\qquad& {\rm SO(9,1)\ vector\ indices}\nonu 
I,J, \ldots = 1, \ldots 8 &\qquad& {\rm SO(8)\ vector\ indices}\nonu
i,j, \ldots = 1, \ldots 4 &\qquad& {\rm SO(4)\ vector\ indices}\nonu
i',j', \ldots = 5, \ldots 8 &\qquad& {\rm SO'(4)\ vector\ indices} 
\eea 
The nonvanishing $PP_{10}$ background fields are given by 
\bea 
ds^2 &=& 2 d x^+ d x^- - \m^2 x^I x^I (dx^+)^2 + dx^I dx^I\nonu
R_{I++J} &=& - \m^2 \d_{IJ} \qquad R_{++} = 8 \m^2\nonu
F_{+1234} &=& F_{+5678} = 4 \m 
\eea 
In light-cone gauge, the $SO(4) \times SO'(4)$ subgroup of the background symmetry group is manifest 
with $x^i$ and $x^{i'}$ transforming as vectors under $SO(4)$ and $SO'(4)$ respectively..  
  Expanding the action  
(\ref{IIBaction}) around this background  to quadratic 
order   
one can organize the fluctuations into (complex) decoupled fields 
$\psi$ which transform in irreducible representations of $SO(4) \times SO'(4)$ \cite{metsaev}.  
We choose our normalizations so that each field $\psi$ contributes a term 
 to the Lagrangian density of the form 
\be 
{\cal L} = {1 \over 4 \k^2} \bar \psi ( \Box - 2 i \mu c \pa_-) \psi 
\label{norm} 
\ee 
where contractions of the $SO(4)\times SO'(4)$ indices are implied where appropriate  
and the bar denotes complex conjugation. 
The operator $\Box = 2 \pa_+ \pa_- + \m^2 x^I x^I \pa_-^2 + \pa_I \pa_I$  
is the scalar Laplacian in \ppten and $c$ is an integer. 
The results needed for the calculation of the D-brane tensions are summarized in table
\ref{pp10table}. 
It displays the fields $\psi$, their definition in  
terms of the original fluctuations (\ref{flucts}), their value of $c$ and  
their irrep of $SO(4) \times SO'(4)$.  
\begin{table}
\begin{center} 
\begin{tabular}{|ccc|c|c|} \hline 
$\psi$ &&   linear combination & c &  irrep\\  
\hline \hline 
$h^\perp_{ij}$ &=& ${1 \over \sqrt{2}} (h^T_{ij} - {1\over 4} \d_{ij}
h^T_{kk})$ &0& ({\bf 9,1})\\ $h^\perp_{i'j'}$ &=& ${1\over \sqrt{2}}
(h^T_{i'j'} + {1\over 4}\d_{i'j'} h^T_{kk})$&0& ({\bf 1,9})\\ $H$&=&
$\half (h^T_{ii}+{i \over 12} a_{ijkl}\e_{ijkl})$&4&({\bf 1,1})\\
$H_{ii'}$&=& $h^T_{ii'} + {i\over 6} a_{i'jkl} \e_{ijkl} $&2&({\bf
4,4})\\
%$\tilde a_{i'j'kl}$&=&${1\over \sqrt{2}} a_{i'j'kl}$&0&\ \\ 
$G_{ij}$&=& ${1 \over
\sqrt{2}}a_{ij}+ {i \over 2 \sqrt{2}} \e_{ijkl} b_{kl} 
$&2&({\bf 6,1})\\ $G_{i'j'}$&=& ${1 \over \sqrt{2}}a_{i'j'}
+{i \over 2 \sqrt{2}}
\e_{i'j'k'l'} b_{k'l'} $&2&({\bf 1,6})\\ &$
b_{ij'}$&&0& ({\bf 4,4})\\ &$ a_{ij'}$&&0& ({\bf 4,4})\\ &
$\f$&&0&({\bf 1,1})\\
\hline 
\end{tabular}
\caption{Decoupled massless fields in \ppten.}
\label{pp10table} \end{center} 
\end{table}
\subsection{Massless propagators in \ppten} 
We will also need the Feynman propagator $G_c(x_1,x_2)$ corresponding to the  
operator $\Box - 2 i \mu c \pa_-$. It can be written as 
\be 
G_c(x_1,x_2) = -i \sum_{\bf n} \int { dp_+ dp_-   
\over (2 \p)^{2}} {e^{i \left( 
p_+ (x^+_1 - x^{+}_2) + p_- (x^-_1 - x^{-}_2) \right)}  
\psi_{\bf n}^{(\m p_-)}(x^I_1)\psi_{\bf n}^{(\m p_-)}(x^I_2) \over  
2 p_+ p_- + \m p_- \sum_I (n_I + \half) + 2 c \m p_- + p_I p_I - i \e} 
\label{momprop} 
\ee 
Here, ${\bf n} = (n_1, n_2,\ldots, n_8)$, and $\psi_{\bf n}^{(\mm)}$ is 
a product of normalized harmonic oscillator 
 eigenfunctions satisfying $(- \pa_I \pa_I + \mm^2 ) \psi_{\bf n}^{(\mm)} =  
2 \mm (\sum_{I} n_I + 1/2)\psi_{\bf n}^{(\mm)}$.  Introducing a Schwinger parameter $s$  
and performing 
the discrete sums and the $p_+,\ p_-$ integrals one gets 
$$G_c(x_1,x_2) = e^{i c \m r^+} G_0(x_1,x_2)$$ 
with 
\be 
G_0(x_1,x_2) = i  \left( {\m r^+ \over \sin \m r^+}\right)^4 \int_0^\infty  
{ds \over (4 \p i s)^5} e^{-{\s + i \e \over 4 i s}} 
\label{proppp10} 
\ee 
with 
\be 
\s = 2 r^+ r^- +  
{\m r^+ \over  \sin \m r^+} \left(  ( x^I_1 x^I_1 + x^I_2 x^I_2 ) \cos \m r^+ 
- 2 x^I_1 x^I_2 \right). 
\ee 
and we have defined $r^\m \equiv x^\m_1 - x^\m_2$. 
The quantity $\s$ is proportional to the invariant distance squared $\F$: 
$$ 
\s = {\m r^+ \over \sin \m \r^+} \F 
$$ 
The integral in (\ref{proppp10}) can be performed to give 
$$ 
G_0(x_1,x_2) = {3  \over 2 \p^5 (\F + i \e)^4} 
$$ 
in agreement with \cite{mathur, gaberdiel}. The limit $\m \to 0$ 
yields the Feynman propagator in Minkowski space. 
In the calculation of D-brane interaction energies, 
we will need the integrated propagator $I^{9-p}_c(r^+, r^-)$   
over $p+1$ longitudinal $x^I$ directions with the transverse $x^I$ 
set to zero. From (\ref{proppp10}) one gets 
\bea 
I_c^{9-p}(r^+, r^-) &=& e^{i c \m r^+} I_0^{9-p}(r^+, r^-)\\ 
&=&{1 \over 4 \p} e^{i c \m r^+}{ (2 \p r^-)^{p-3} 
\m ^{3-p} \over  \sin^4 \m r^+}\G (3 -p). 
\label{intproppp10} 
\eea

\subsection{Massless modes in \ppsix} 
\label{pp6modes} 
We use the following index conventions: 
\bea 
\m,\n,\ldots = 0, \ldots,9 &\qquad& {\rm SO(9,1)\ vector\ indices}\nonu
I,J, \ldots = 1, \ldots 8 &\qquad& {\rm SO(8)\ vector\ indices}\nonu
i,j, \ldots = 1,2 &\qquad& {\rm U(1)\ vector\ indices}\nonu
i',j', \ldots = 3,4 &\qquad& {\rm U'(1)\ vector\ indices}\nonu
a,b, \ldots = 5, \ldots ,8 &\qquad& {\rm SO(4)\ vector\ indices} 
\eea 
It is convenient to work with complex coordinates $z,w$ 
instead of $x^i$ and $x^{i'}$ : 
\bea 
z &=& x^1 + i x^2\\ 
w &=& x^3 + i x^4 
\eea 
The nonvanishing \ppsix background fields are then given by 
\bea 
ds^2 &=& 2 d x^+ d x^- - \m^2 (z \bar z + w \bar w)(dx^+)^2 + dz d\bar z 
+ dw d\bar w + dx^a dx^a\nonu 
R_{z++\bar z} &=& - \half \m^2 \qquad R_{w++\bar w} = - \half \m^2  
 \qquad R_{++} = 4 \m^2\nonu
F_{+z \bar z} &=& F_{+w \bar w} = i \m 
\eea  
In light-cone gauge, the $U(1)\times U'(1)\times SO(4)$ subgroup of
the background symmetry group is manifest with $z$ and $w$ carrying
charge -1 under $U(1)$ and $U'(1)$ respectively and $x^a$ transforming
as a vector under $SO(4)$.  We can again organize the massless modes
into decoupled fields $\psi$ which transform in irreps of $U(1)\times
U'(1)\times SO(4)$ and whose contribution to the action is
characterized by an integer $c$ as in (\ref{norm}), where the scalar
Laplacian is now $\Box = 2 \pa_+ \pa_- + \m^2 (z\bar z + w \bar w)
\pa_-^2 + 4 \pa_z \pa_{\bar z} + 4 \pa_w \pa_{\bar w}+ \pa_a
\pa_a$. The results of this heartwarming calculation are summarized in
table \ref{pp6table}. It displays the fields $\psi$, their definition
in terms of the original fluctuations (\ref{flucts}), their value of
$c$ and their irrep of $U(1)\times U'(1)\times SO(4)$. We use the
notation ${\bf d_{(q,q')}}$ for the $d$-dimensional representation of
$SO(4)$ with charges $(q,q')$ under $U(1)\times U'(1)$.
\\ 
\begin{table}
\begin{center} 
\begin{tabular}{|ccc|c|c|} \hline 
$\psi$ &&   linear combination & c &  irrep\\  
\hline \hline 
$\tilde h_{zz}$ &=& $2 h^T_{zz}$&0& ${\bf 1_{(2,0)}}$\\ 
$\tilde h_{ww}$ &=& $2 h^T_{ww}$&0& ${\bf 1_{(2,0)}}$\\ 
$h^\perp_{ab}$ &=& ${1\over \sqrt{2}} (h^T_{ab} +  \d_{ab} (h^T_{z\bar z}+ h^T_{w \bar w}))$&0& 
  ${\bf 9_{(0,0)}}$\\ 
$\tilde a_{ab}$ &=& ${1 \over \sqrt{2}} a_{ab}$ &0& ${\bf 6_{(0,0)}}$ \\ 
$H^\pm_{zw}$ &=& $2( h^T_{zw} \pm a_{zw} )$&0&${\bf 1_{(1,1)}}$\\ 
%$H^-_{zw}$ &=& $2( h^T_{zw} - a_{zw} )$&0&${\bf 1_{(1,1)}}$\\ 
$H^\pm_{z\bar w}$ &=& $2( h^T_{z\bar w} \pm a_{z\bar w} )$&$\pm 2$&${\bf 1_{(1,-1)}}$\\ 
%$H^-_{z\bar w}$ &=& $2( h^T_{z\bar w} - a_{z\bar w} )$&2&${\bf 1_{(1,-1)}}$\\ 
$H_0$&=& $\sqrt{2} (h^T_{z \bar z} + h^T_{w \bar w}) + {1 \over \sqrt{2}} \f$&0&${\bf 1_{(0,0)}}$\\ 
$H$&=& $2 h^T_{z \bar z} - \half \f +2 a_{z \bar z}$ &2&${\bf 1_{(0,0)}}$\\ 
$H'$&=& $2 h^T_{w \bar w} - \half \f +2 a_{w \bar w}$ &2&${\bf 1_{(0,0)}}$\\ 
$H^\pm_{az}$&=& $\sqrt{2}(h^T_{az} \pm a_{az})   $&$\mp 1$&${\bf 4_{(1,0)}}$\\ 
%$H^-_{az}$&=& $\sqrt{2}(h^T_{az} - a_{az})   $& 1&${\bf 4_{(1,0)}}$\\ 
$H^\pm_{aw}$&=& $\sqrt{2}(h^T_{aw} \pm  a_{aw})   $&$\mp 1$&${\bf 4_{(0,1)}}$\\ 
%$H^-_{aw}$&=& $\sqrt{2}(h^T_{aw} - a_{aw})   $& 1&${\bf 4_{(0,1)}}$\\ 
$G_0$&=& ${1 \over \sqrt{2}} (a - 4 a_{z \bar z w \bar w})$&0&${\bf 1_{(0,0)}}$\\ 
$G_0'$&=& $- i \sqrt{2}  (b_{z \bar z} - b_{w \bar w})$&0&${\bf 1_{(0,0)}}$\\ 
$G$&=& $ \sqrt{2}  (b_{z \bar z} + b_{w \bar w})  
+{1 \over \sqrt{2}} (a + 4 a_{z \bar z w \bar w})$&2&${\bf 1_{(0,0)}}$\\ 
\hline 
\end{tabular} 
\caption{Decoupled massless fields in \ppsix.}
\label{pp6table} 
\end{center}
\end{table}
\subsection{Massless propagators in \ppsix} 
The Feynman propagator $G_c(x_1,x_2)$ corresponding to the  
operator $\Box - 2 i \mu c \pa_-$ can be written as 
\be 
-i \sum_{\bf n} \int { dp_+ dp_- d^4 p_a  
\over (2 \p)^6} {e^{i \left( 
p_+ (x^+_1 - x^{+}_2) + p_- (x^-_1 - x^{-}_2) + p_a (x^a_1 -  x^a_2) 
\right)} \psi_{\bf n}^{(\m p_-)}(x^A_1)\psi_{\bf n}^{(\m p_-)}(x^A_2) \over  
2 p_+ p_- + \m p_- \sum_A (n_A + \half) + 2 c \m p_- + p_a p_a - i \e} 
\label{momproppp6} 
\ee 
Here, $A=1 \ldots 4$,  ${\bf n} = (n_1, n_2,n_3, n_4)$ and $\psi_{\bf n}^{(\mm)}$ is 
a product of harmonic oscillator 
 eigenfunctions satisfying $(- \pa_i \pa_i - \pa_{i'}\pa_{i'} + \mm^2 ) \psi_{\bf n}^{(\mm)} =  
2 \mm (\sum_{A} n_A + 1/2)\psi_{\bf n}^{(\mm)}$. 
Introducing a Schwinger parameter $s$  
and performing 
the discrete sums and the $p_+,\ p_-,\ p_a$ integrals one gets 
$$G_c(x_1,x_2) = e^{i c \m r^+} G_0(x_1,x_2)$$ 
with 
\be 
G_0(x_1,x_2) = i  \left( {\m r^+ \over \sin \m r^+}\right)^2 \int_0^\infty  
{ds \over (4 \p i s)^5} e^{ -{\s + i \e \over 4 i s}} 
\label{proppp6} 
\ee 
with 
\bea 
\s &=&  2 r^+ r^- +  r^a r^a  \nonu 
&+& 
{\m r^+ \over  \sin \m r^+} \left(  ( x^i_1 x^i_1 + x^{i'}_1 x^{i'}_1 +  
x^i_2 x^i_2 + x^{i'}_2 x^{i'}_2) \cos \m r^+ 
- 2 (x^i_1 x^i_2 + x^{i'}_1 x^{i'}_2) \right) \nonumber
\eea 
and we have defined $r^\m \equiv x^\m_1 - x^\m_2$. 
In the calculation of D-brane interaction energies, we need the 
integrated propagator, denoted by  $I^{9-p}_c$,   
over $p+1$ longitudinal D-brane directions  
(which we take be a subset of the pp-wave directions  
$x^1, \ldots x^4$) and with the remaining pp-wave coordinates 
set to zero: 
\bea 
I_c^{9-p}(r^+, r^-, r^a) &=& e^{i c \m r^+} I_0^{9-p}(r^+, r^-, r^a)\\ 
&=& {1 \over 4 \p} e^{i c \m r^+}{\left(\p r^2\right)^{p-3} 
(\m r^+)^{1-p} \over \sin^2 \m r^+}\G (3 -p) 
\label{intproppp6} 
\eea 
where $r^2 \equiv 2 r^+ r^- + r^a r^a $.

\end{appendix} 
 
\end{document}